\documentclass[11pt]{article}
\usepackage{a4}
\usepackage{amsmath}
\usepackage{amssymb}
\usepackage[dvips]{graphicx}
\usepackage{subfigure}
\def\beq{\begin{equation}}
\def\eeq{\end{equation}}
\def\beqa{\begin{eqnarray}}
\def\eeqa{\end{eqnarray}}
\def\n{\nonumber \\}

\def\e{{\,\rm e}\,}

\newcommand {\tr}{{\rm tr}\,}

\newcommand {\CTr}{{\cal T}r\,}

\def\dag{\dagger}
\newcommand{\id}{{1\!\!1}} %% identity matrix

\begin{document}

\vspace*{1.0cm}
\begin{flushright}
{SAGA-HE-264}
\end{flushright}
\vskip 1.0cm

\begin{center}
{\large{\bf Chiral Fermions and the Standard Model\\ 
from the Matrix Model Compactified on a Torus}}
\vskip 1.0cm

{\large Hajime Aoki\footnote{e-mail
 address: haoki@cc.saga-u.ac.jp}
}
\vskip 0.5cm

{\it Department of Physics, Saga University, Saga 840-8502,
Japan  }\\

\end{center}

\vskip 1cm
\begin{center}
\begin{bf}
Abstract
\end{bf}
\end{center}
It is shown that the IIB matrix model compactified on a six-dimensional torus 
with a nontrivial topology can provide
chiral fermions and matter content close to the standard model
on our four-dimensional spacetime.
In particular, generation number three is given by the
Dirac index on the torus.

%%%%%%%%%%%%%%%%%%%%%%%%%%%%%%%%%%%%%%%%%%%%%%%% 
\newpage
\setcounter{footnote}{0}
\section{Introduction}
\setcounter{equation}{0}

Matrix models are a promising candidate to 
formulate the superstring theory nonperturbatively \cite{Banks:1996vh,IKKT},
and they indeed include quantum gravity and gauge theory.
One of the important subjects in such studies is to 
connect these models to phenomenology. 
Spacetime structures can be analyzed dynamically 
in the IIB matrix model \cite{Aoki:1998vn},
and four dimensionality seems to be preferred \cite{Aoki:1998vn,Nishimura:2001sx}.
Assuming that four-dimensional spacetime is obtained, 
we next want to show the standard model of
particle physics on it.
An important ingredient of the standard model is the chirality of fermions.
Chirality also ensures the existence of massless fermions,
since, otherwise, quantum corrections would induce mass
of the order of the Planck scale or of the Kaluza-Klein scale in general.

A way to obtain chiral spectrum in our spacetime 
is to consider topologically nontrivial configurations 
in the extra dimensions\footnote{Having
this mechanism in mind, we analyzed the dynamics of a model on a fuzzy 2-sphere
and showed that topologically nontrivial configurations are indeed realized  \cite{AIMN}.
Models of four-dimensional field theory with fuzzy extra dimensions were
studied in
\cite{Steinacker:2007ay}.}.
Owing to the index theorem \cite{Atiyah:1971rm},
the topological charge of the background 
provides the index of the Dirac operator,
{\it i.e.}, 
the difference in the numbers of chiral zero modes,
which then produce massless chiral fermions on our spacetime.
Generalizations of the index theorem 
to matrix models or noncommutative (NC) spaces 
with finite degrees of freedom were provided
by using a Ginsparg-Wilson (GW) relation\footnote{GW Dirac operators
on a fuzzy 2-sphere and a NC torus were given 
in \cite{balagovi} and \cite{Nishimura:2001dq}, respectively.
A general formulation for constructing GW Dirac operators on general geometries
and defining the corresponding index theorem was provided in \cite{AIN2}.}
developed in the lattice gauge theory \cite{GinspargWilson}.

In $M^4 \times S^2 \times S^2$ embeddings in the IIB matrix model, however,
we could not obtain a chiral spectrum on $M^4$,
even though the IIB matrix model is chiral in ten dimensions,
and topological configurations give chiral zero modes on  $S^2 \times S^2$,
since the remainder dimensions $M^{10}/(M^4 \times S^2 \times S^2)$ interrupt \cite{Aoki:2010hx}.
This obstacle arises generally in the cases with remainder dimensions,
such as the coset space constructions.
We thus have to consider the situations where topological configurations 
are embedded in the entire six extra dimensions\footnote{In
the case of spheres,
if we also embed topological structures in the
direction of the thickness of the sphere shell,
the problem is resolved.}.

We then consider compactifications on tori, such as $M^4 \times T^6$.
Toroidal compactifications in the matrix models were studied in
\cite{Taylor:1996ik, Connes:1997cr},
and their unitary matrix formulations were also considered 
\cite{Polychronakos:1997fw}.
Moreover, a formulation for gauge theories with adjoint matter
in nontrivial topological sectors on a NC torus
was given by using the Morita equivalence \cite{Ambjorn:1999ts}.
For the fundamental matter, since the Morita equivalence is not satisfied in this case,
a matrix model formulation was provided in a purely algebraic way
\cite{Aoki:2008ik}\footnote{
All the formulations for toroidal compactifications 
correspond to imposing the periodic or the 
twisted boundary conditions
on the matrices, rather than embedding manifolds in larger-dimensional spaces.
In this sense, they are related to orbifolds and orientifolds.
Their matrix model formulations were studied, for instance, in
\cite{Aoki:2002jt} and \cite{Itoyama:1997gm}, respectively.
}.

In this paper, we begin with 
a gauge theory with adjoint matter in the trivial topological sector,
since adjoint matter naturally arises from the matrix models whose action
is written by the commutators.
We then introduce block-diagonal matrix configurations
as topologically nontrivial  gauge field backgrounds.
The off-diagonal blocks of the adjoint matter field,
which are in the bifundamental representations of the gauge group
produced by the background,
thus obtain nonzero Dirac indices.
Note that nontrivial topologies are given by the 
backgrounds, not by imposing suitable 
boundary conditions by hand.
We further show that such configurations, 
when considered in the extra dimensions in the IIB matrix model,
indeed give chiral spectrum on our spacetime.
We also study the dynamics of these configurations by
investigating their classical actions, 
and find that they appear in the continuum limit
as in the gauge theories on the commutative spaces.
We finally present an example of a configuration that
gives matter content close to the standard model\footnote{
Almost all the arguments and results presented in this paper
are valid in general contexts with toroidal compactifications
and nontrivial topologies, and do not depend on our specific 
settings, {\it i.e.}, the unitary matrix formulation and the NC space.
Here, we exploit the unitary matrix formulation since it is described by
finite matrices.
We also think that noncommutativities arise naturally  
if we start from the matrix models
\cite{Connes:1997cr, Aoki:1999vr}.
We will also discuss in section \ref{sec:conclusion} that
the noncommutativity may give a seed to select 
matrix configurations with three generations
dynamically from many possible classical solutions.
}.

In section \ref{sec:actions}, we briefly review the finite matrix formulation of 
gauge theories with adjoint matter on a NC torus, including
the formulation of the GW Dirac operator and the index theorem.
Then in section \ref{sec:top_con},
we introduce block-diagonal configurations
as topological backgrounds.
Explicit forms of the configurations on two-dimensional and six-dimensional tori are given in 
section \ref{sec:2dtorus} and section \ref{sec:6dtorus}, respectively.
Dynamics of the configurations are studied in section \ref{sec:cla_act}.
In section \ref{sec:SMembedding},
we show an example of a configuration that
gives matter content close to the standard model.
Section \ref{sec:conclusion} is devoted to conclusions and discussion.
In appendix \ref{sec:cal_index},
we calculate the index of the GW Dirac operator.

%%%%%%%%%%%%%%%%%%%%%%%%%%%%%%%%%%%%%%%%%%%%%%%%%%%%%%%%%%% 
\section{Gauge theory with adjoint matter on a NC torus}
\label{sec:actions}
\setcounter{equation}{0}

In this section, we briefly review the finite matrix formulation of 
gauge theories with adjoint matter on a noncommutative (NC) torus.
For details, see \cite{Ambjorn:1999ts}, for instance.
Here, we consider a simple setting that gives a topologically trivial sector,
however.

An action for the gauge fields on a $d$-dimensional NC torus
can be given by the twisted Eguchi-Kawai model 
\cite{Eguchi:1982nm,GonzalezArroyo:1982ub}
\beq
S_{b} = -{\cal N} \beta \, \sum_{\mu \ne \nu} 
{\cal  Z}_{\nu\mu}
\tr ~\Bigl(V_\mu\,V_\nu\,V_\mu^\dag\,V_\nu^\dag\Bigr) 
 + d (d-1)\beta {\cal N}^2 \ ,
\label{TEK-action}
\eeq
with $\mu,\nu=1,\ldots ,d$.
Here, $V_\mu$ denote $U({\cal N})$ matrices representing the link variables on the lattice,
$\beta$ stands for the lattice gauge coupling constant,
and ${\cal  Z}_{\nu\mu}$ are $Z_{\cal  N}$ factors that are assumed to be specified
to give the topologically trivial sector.
The constant term is added to 
make the action vanish at its minimum.

Actions for adjoint matter are given by using covariant 
forward and backward difference operators 
\beqa
\nabla_\mu \psi&=&
%\frac{1}{\epsilon}\left[\hat{U}_\mu 
%(\Gamma^{(n)}_\mu \psi \Gamma_\mu^{(N)\dag})
%- \psi  \right]
%= 
\frac{1}{\epsilon}\left(V_\mu \, \psi \, V_\mu ^{\dag}
- \psi  \right) \ , \n
\nabla_\mu^* \psi &=&
%\frac{1}{\epsilon}\left[\psi - 
%(\Gamma_\mu^{(n)\dagger} \hat{U}_\mu^{\dagger}\Gamma^{(n)}_\mu) 
%(\Gamma_\mu^{(n)\dagger} \psi  \Gamma^{(N)}_\mu) \right] 
%=
\frac{1}{\epsilon}\left(\psi - V_\mu ^{\dagger} \,
\psi \,  V_\mu \right)  \  ,
\label{def-cov-shift}
\eeqa
with $V_\mu \in U({\cal N})$ introduced above.
$\epsilon$ is an analog of the lattice spacing.
For instance, a Wilson-Dirac operator $D_{\rm W}$ is defined as
\beq
D_{\rm W}=\frac{1}{2}\sum_{\mu=1}^d
\left\{\gamma_\mu\left(\nabla_\mu^* 
+\nabla_\mu \right) - \epsilon  \nabla_\mu^* \nabla_\mu \right\} \ ,
\label{def-Wilson-Dirac}
\eeq
where $\gamma_\mu$ are $d$-dimensional Dirac matrices.

One can also define a Ginsparg-Wilson (GW) Dirac operator as\footnote{
We explain it according to the general formulation \cite{AIN2} here,
while it was obtained by applying the Neuberger's overlap Dirac operator
to a NC torus \cite{Nishimura:2001dq}.}
\beq
D_{\rm GW}=\frac{1}{\epsilon} (1 - \gamma \hat{\gamma}) \ ,
\label{def-GW-Dirac}
\eeq
where $\gamma$ 
is an ordinary chirality operator on the $d$-dimensional space,
and $\hat{\gamma}$ is a modified one defined as
\beqa
\hat\gamma &=& \frac{H}{\sqrt{H^2}} \ , 
\label{def-gamma-hat}\\
H &=& \gamma \left(1- \epsilon D_{\rm W}\right) \ ,
\label{H-def}
\eeqa
with $D_{\rm W}$ given in (\ref{def-Wilson-Dirac}).
They satisfy the relations
\beq
\gamma^\dagger=\gamma~,~~
\hat\gamma^\dagger=\hat\gamma~,~~
\gamma^2=\hat\gamma^2=1 \ .
\eeq
Then, 
by the definition (\ref{def-GW-Dirac}), the Dirac operator
satisfies a GW relation
\beq
\gamma D_{\rm GW} + D_{\rm GW} \hat\gamma =0 \ .
\eeq
Hence, the index, {\it i.e.}, the difference in the numbers 
of chiral zero modes, is given by
the trace of the chirality operators as
\beq
{\rm{index}}(D_{\rm GW})
=\frac{1}{2}\CTr[\gamma +\hat{\gamma}] \ ,
\label{ITtrivial}
\eeq
where $\CTr$ is the trace over the whole configuration space.
Since the definition of $\hat\gamma$ depends on the link variables $V_\mu$, 
the right-hand side (rhs) of (\ref{ITtrivial}) is a functional of the
gauge field configurations.
It also takes only integer values,
since it is a trace of sign operators.
Moreover, it is shown to become the Chern character with star product 
in the continuum limit for the fundamental matter \cite{Iso:2002jc}.
It then gives a noncommutative
generalization of the topological charge for the gauge field backgrounds.
Thus, eq.~(\ref{ITtrivial}) gives
an index theorem on the NC torus.

We expect, however, that the rhs of (\ref{ITtrivial})
 vanishes for any configurations $V_\mu$
that survive in the continuum limit
because of the following reasons:
First, the rhs of (\ref{ITtrivial}) is considered to have an appropriate continuum limit,
as shown for the fundamental matter case in \cite{Iso:2002jc}.
Since the adjoint matter is chiral-anomaly-free in 2 (mod 4) dimensions,
it must vanish.
Second, since we now begin with the matrix model (\ref{TEK-action})
describing the trivial module,
only the topologically trivial sector
appears in the continuum limit, 
as shown in \cite{Aoki:2006sb, Aoki:2006zi}.
We therefore need some modifications in order to have 
nontrivial topologies, which we will study in the next section.

%%%%%%%%%%%%%%%%%%%%%%%%%%%%%%%%%%%%%%%%%%%%%%%%%%%%%%%%
\section{Topological configurations}
\label{sec:top_con}
\setcounter{equation}{0}

As topologically nontrivial gauge configurations,
we introduce the following block-diagonal matrices:
\begin{equation}
V_\mu=
\begin{pmatrix}
 V_\mu^1 & &&\cr
& V_\mu^2 &&\cr
&& \ddots& \cr
&&& V_\mu^h
\end{pmatrix} 
\label{V_mu_block} \ , 
\end{equation}
with $h$ blocks and $\mu=1,\ldots,d$.
As we will see in the following sections,
each block produces gauge group $U(p^a)$
with $a=1,\ldots,h$.

We also introduce the following projection operators $P^a$ with $a=1,\ldots ,h$, which
pick up the space that $a$th block acts:
\beq
P^a=
\begin{pmatrix}
 \ddots &&&&\cr
& 0 &&&\cr
&& \id && \cr
&&& 0& \cr
&&&& \ddots
\end{pmatrix} \ .
\eeq
Since $P^a$ commutes with the chirality operator (\ref{def-gamma-hat}) 
and the Dirac operator (\ref{def-GW-Dirac}),
the index theorem (\ref{ITtrivial}) is satisfied in each space projected
by $P^{a}$ as
\begin{equation}
{\rm{index}}(P^{aL}P^{bR}D_{\rm GW})
=\frac{1}{2}\CTr[P^{aL}P^{aR}(\gamma +\hat{\gamma})] \ , 
\label{ITprojected}
\end{equation}
where the superscript $L$ ($R$) means that 
the operator acts from the left (right) on matrices:
${\cal O}^LM \equiv {\cal O}M,~{\cal O}^RM \equiv M {\cal O}$.
$P^{aL}P^{bR}$ picks up the following block $\psi^{ab}$ from 
the matter field $\psi$
in the adjoint representation:
\beq
\psi=
\begin{pmatrix}
\psi^{11} & \psi^{12} & \cdots & \psi^{1h} \cr
\psi^{21} & \psi^{22} & \cdots & \psi^{2h} \cr
\vdots & \vdots & \ddots & \vdots \cr
\psi^{h1}& \psi^{h2}& \cdots & \psi^{hh}
\end{pmatrix} \ ,
\label{psi_block_decompose}
\eeq
where we decompose $\psi$ into blocks in the same way as (\ref{V_mu_block}).
The diagonal blocks $\psi^{aa}$ are in the adjoint representations
under the gauge group,
while the off-diagonal blocks $\psi^{ab}$ with $a \neq b$ are in the
bifundamental representations.
As shown in the following sections, the index of each block 
(\ref{ITprojected})
can have nonzero values,
although the total matrix $\psi$ has a vanishing index.

In the remainder of this section, we show that, 
by considering the configurations (\ref{V_mu_block}) 
with $d=6$ in the extra dimensions in the IIB matrix model,
chiral fermions on our four-dimensional spacetime are obtained.
See \cite{Aoki:2010hx} for detailed arguments.
For $d=2$ (mod $4$),
the topological charge becomes the $(d/2)$th Chern character, with $d/2$ 
being an odd integer.
Hence, $\psi^{ab}$ and $\psi^{ba}$, which are in the conjugate 
representations under the
gauge group, have the opposite indices. 
We denote the corresponding chiral zero modes as $\psi^{ab}_R$ and $\psi^{ba}_L$,
where the subscripts $R$ and $L$ stand for the chirality.
(Choosing $\psi^{ab}_L$ and $\psi^{ba}_R$ instead would give the identical
results shown below.)
Taking spinors $\varphi$ on our four-dimensional spacetime as well,
we obtain the following possible Weyl spinors:
\beqa
\varphi_R \otimes \psi^{ab}_R \ , \label{d2RR}\\
\varphi_L \otimes \psi^{ba}_L \ , \label{d2LL}\\
\varphi_L \otimes \psi^{ab}_R \ , \label{d2LR}\\
\varphi_R \otimes \psi^{ba}_L \ .\label{d2RL}
\eeqa
The spinors (\ref{d2RR}) and (\ref{d2LL}) are in the charge conjugate representations 
to each other under the gauge and the Lorentz groups;
so are (\ref{d2LR}) and (\ref{d2RL}).

Since the IIB matrix model has a ten-dimensional Majorana-Weyl spinor,
we now impose these conditions.
By the Weyl condition, (\ref{d2RR}) and (\ref{d2LL})
 are chosen.
(Choosing (\ref{d2LR}) and (\ref{d2RL}) gives identical results.)
Since the four-dimensional Weyl spinors
$\varphi_R$ in (\ref{d2RR}) and $\varphi_L$ in (\ref{d2LL})
are in the different representations under the gauge group,
they give chiral spectrum on our spacetime,
although we still have a doubling of (\ref{d2RR}) and (\ref{d2LL}).
Furthermore,
by the Majorana condition, (\ref{d2RR}) and (\ref{d2LL})
are identified.
(So are (\ref{d2LR}) and (\ref{d2RL}).)
Then, the unwanted doubling of (\ref{d2RR}) and (\ref{d2LL})
is also resolved.

%%%%%%%%%%%%%%%%%%%%%%%%%%%%%%%%%%%%%%%%%%%%%%%%%%%%%%%
\section{Two-dimensional torus}
\label{sec:2dtorus}
\setcounter{equation}{0}

In this section, we show explicit forms of the configurations 
(\ref{V_mu_block}) with $d=2$.
In the context of $M^4  \times T^6$ compactifications in the IIB matrix model,
this $T^2$ corresponds to the one in $T^6=T^2\times T^2\times T^2$.

We consider the following configurations:
\beq
V_\mu = 
\begin{pmatrix}
 \Gamma_\mu^1 \otimes \id_{p^1}& &&\cr
& \Gamma_\mu^2 \otimes \id_{p^2} &&\cr
&& \ddots& \cr
&&& \Gamma_\mu^h \otimes \id_{p^h}
\end{pmatrix} 
\label{2d_Vmu_exp} \ , 
\eeq
with $\mu =1,2$.
The factors $\id_{p^a}$ with $a=1,\ldots ,h$ give
gauge group $U(p^1) \times \cdots \times U(p^h)$.
The matrices $\Gamma_\mu^a$ represent NC tori with
magnetic fluxes specified by integers $q^a$.
The configurations (\ref{2d_Vmu_exp})
are classical solutions for the action (\ref{TEK-action}),
as shown in \cite{Aoki:2006sb}.

We now show some details about formulations of a NC torus.
For more details, see ref. \cite{Aoki:2008ik}.
We use the same conventions as in \cite{Aoki:2008ik} here.
The matrix $\Gamma_\mu^a$  is a shift operator
on a dual torus specified by a set of integers
$n^a, m^a, j^a, k'^a$ for each $a$.
They satisfy the Diophantine equation,
\beq
m^a j^a + n^a k'^a =1 \ .
\eeq
We also introduce an original torus specified by a set of integers
$N, s, r, k$, satisfying the  Diophantine equation,
\beq
2rs -kN = -1 \ .
\label{Dio_rskN}
\eeq
The dual torus and the original torus are related by the integer $q^a$,
which specifies the magnetic flux on the dual torus, as\footnote{In 
\cite{Aoki:2008ik}, the dual torus is determined by the two integers $p$ and $q$,
which specify the gauge group $U(p)$ and the abelian flux.
The present case corresponds to $p=p^a, q=p^a q^a$, and hence,
$p_0=p^a, \tilde{p}=1, \tilde{q}=q^a$. }
\beq
m^a = -s + k q^a \ ,~~~
n^a = N - 2r q^a \ .
\label{rel_mn_1q}
\eeq
Equation (\ref{rel_mn_1q}) can be inverted as
\beq
1=2r m^a + k n^a \ ,~~~
q^a = N m^a + s n^a \ .
\label{rel_1q_mn}
\eeq

Explicit forms of the coordinate and the shift operators on the dual torus
are given, for instance, as
\beqa
Z_1^{a} = W_{n^a}  ~ &,& ~~  Z_2^{a}= (V_{n^a})^{j^a} \ , \n
\Gamma_1^{a} = V_{n^a} ~ &,& ~~ \Gamma_2^{a} = (W_{n^a})^{-m^a} \ ,
\label{2d_ZG_exp}
\eeqa
in terms of the shift and  clock matrices
\beq
V_{n}=\begin{pmatrix}
0&1& & &0\cr &0&1& & \cr & &\ddots&\ddots& \cr
 & & &\ddots&1\cr1& & & &0\cr
\end{pmatrix}  \ , \ \
W_{n}=\begin{pmatrix}
1& & & & \cr
&e^{2\pi i/n}& & &\cr & &\e^{4\pi i/n}& & \cr
 & & &\ddots& \cr & & & &\e^{2\pi i(n-1)/n}\cr
 \end{pmatrix} \ , 
\label{def_shit_clock}
\eeq
which are $U(n)$ matrices 
obeying the commutation relations
\beq
V_n W_n=
\e^{2\pi i/n}\,W_nV_n \ .
\label{alg_VW}
\eeq

The off-diagonal block $\psi^{ab}$ in (\ref{psi_block_decompose})
can be interpreted as in the fundamental representation,
if we identify the $b$th block as an original torus.
The corresponding integer $q$ is thus given by
(\ref{rel_1q_mn}),
with $N$ and $s$ replaced by $n^b$ and $-m^b$, respectively.
Substituting (\ref{rel_mn_1q}) and using (\ref{Dio_rskN}), we obtain
\beq
%q^{ab}= 
n^b m^a-m^b n^a=q^a-q^b \ .
\label{qab_nm_qaqb}
\eeq
Then, the index for the block $\psi^{ab}$ (\ref{ITprojected})
should become
\beq
\frac{1}{2}\CTr[P^{aL}P^{aR}(\gamma+\hat{\gamma})] =p^ap^b(q^a-q^b) \ .
\label{2d_index_q_ab}
\eeq
Indeed, as shown by the explicit calculations in appendix \ref{sec:cal_index},
eq.~(\ref{2d_index_q_ab}) is satisfied 
in general, except for the rare cases
with $|r|=1$, $n^a=1$, and $n^b=2|q^a-q^b|+1$,
or the cases with $n^a$ and $n^b$ reversed.
As long as we consider the cases with the block sizes $n^a$ greater than one,
eq.~(\ref{2d_index_q_ab}) is satisfied.
The Monte Carlo results in \cite{Aoki:2009fs}
also support (\ref{2d_index_q_ab}).
Equation (\ref{2d_index_q_ab}) means that the index of 
each component 
in the $(p^a,\bar{p^b})$ representation  
under the gauge group $U(p^a) \times U(p^b)$
is $q^a-q^b$.
By using a relation 
\beq
n^a-n^b=-2r(q^a-q^b)
\label{rel_nanbqaqb}
\eeq
given by (\ref{rel_mn_1q}),
eq.~(\ref{2d_index_q_ab}) is rewritten as 
\beq
\frac{1}{2}\CTr[P^{aL}P^{aR}(\gamma+\hat{\gamma})] =-\frac{1}{2r}p^ap^b(n^a-n^b) \ .
\eeq
The same equation was given for the fuzzy 2-sphere case 
in eq.~(5.4) of \cite{Aoki:2010hx}\footnote{The case with 
the fundamental matter was studied in \cite{AIN3}.
The formulation was further extended to 
$S^2\times S^2$ in \cite{Aoki:2009cv}.
},
except for the factor $2r$.

\subsection{Classical actions}
\label{sec:cla_act}

We now study the dynamics of the configurations (\ref{2d_Vmu_exp})
by evaluating their classical actions (\ref{TEK-action}).
Similar analyses were given in \cite{Aoki:2006sb},
but the present case corresponds to the situation
where all the configurations  are in the topologically trivial sector 
in the sense of \cite{Aoki:2006sb}, where the topology was defined 
in terms of the total matrix.
Now,
the nontrivial topologies arise from the blocks, as explained in 
section \ref{sec:top_con}.

We take $p^1=\cdots =p^h=1$ without loss of generality.
We also choose
the integers $r$ and $k$ specifying the original torus to be $r=-1$, $k=-1$, 
which give $s=\frac{N+1}{2}$ from (\ref{Dio_rskN}),
following the previous works 
\cite{Aoki:2006sb, Aoki:2006zi, Aoki:2009fs}.
From (\ref{rel_mn_1q}),
$n^a=N+2q^a$ and $m^a=-\frac{n^a+1}{2}$ are 
determined{\footnote{Actually, this fixing of $r$ and $k$ is not necessary 
in the following analysis, since, from 
(\ref{rel_mn_1q}) and (\ref{rel_1q_mn}),
one can obtain a relation
\[\frac{s}{N}+\frac{m^a}{n^a}
=\frac{q^a}{Nn^a}
=-\frac{1}{2r}\left(\frac{1}{N}-\frac{1}{n^a}\right) \ .
\]}.
It then follows from (\ref{2d_ZG_exp}) that
\beq
\Gamma_1^a \Gamma_2^a = \e^{2\pi i\frac{n^a+1}{2n^a}}\Gamma_2^a \Gamma_1^a \ .
\eeq
Choosing the phase ${\cal  Z}_{\mu\nu}$in the action (\ref{TEK-action}) as
\beq
{\cal  Z}_{12}=\e^{2\pi i\frac{{\cal N}+1}{2{\cal N}}} \ ,
\eeq
the actions for the configurations (\ref{2d_Vmu_exp}) become
\beq
S = -2{\cal N}\beta \sum_{a=1}^h n^a 
\cos\left(\pi\left(\frac{1}{{\cal N}}-\frac{1}{n^a}\right)\right) +2\beta{\cal N}^2 \ .
\label{cla_act_2d}
\eeq
For $h$ blocks with the same sizes, $n^1=\ldots =n^h$, 
(\ref{cla_act_2d}) becomes
\beq
S^{h}=\beta \pi^2 (h-1)^2 -\frac{1}{12}\beta \pi^4 (h-1)^4\frac{1}{{\cal N}^2}
+{\cal O}\left(\left(1/{\cal N}\right)^4\right) \ .
\eeq

We now study the cases where the block sizes are different.
For simplicity, we consider the cases with $h=3$ and
the size of the total matrix
${\cal N}$ and that of the third block $n^3$ fixed.
They correspond to the cases where we focus on the two blocks
with the other $h-2$ blocks fixed.
The action (\ref{cla_act_2d}) for $n \equiv n^1$ becomes
\beq
S(n) = -2{\cal N}\beta \left[ 
n \cos\left(\pi\left(\frac{1}{{\cal N}}-\frac{1}{n}\right)\right) 
+({\cal N}-n^3-n) \cos\left(\pi\left(\frac{1}{{\cal N}}
-\frac{1}{{\cal N}-n^3-n}\right)\right)
\right] \ ,
\label{clac_diff_size}
\eeq
where we did not write the constant terms.
As shown in figure \ref{fig:claact},
$S(n)$ has its minimum at the middle point 
$n=\frac{{\cal N}-n^3}{2}$
with a flat plateau around it.
\begin{figure}
\begin{center}
\includegraphics[height=4cm]{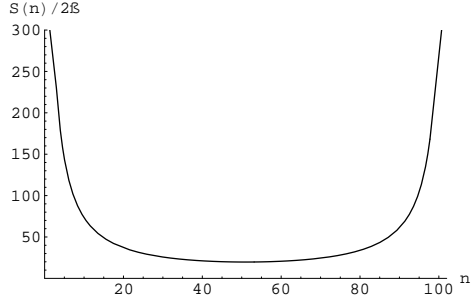}
 \caption{The classical action (\ref{clac_diff_size}) as a function of $n$
 is displayed.
 Here, we take ${\cal N}=153$ and $n^3=51$.}
\label{fig:claact}
\end{center}
\end{figure}
The function $S(n)$ is in fact symmetric at the middle point
and convex downwards.
We note that the middle point corresponds to the configuration
where the first and second blocks have the same size,
which gives trivial topology to the off-diagonal block $\psi^{12}$.
We then consider the difference in the actions between
the topologically trivial and nontrivial configurations.
By expanding in $1/({\cal N}-n^3)$, we obtain
\beq
S\left(\frac{{\cal N}-n^3}{2} +m \right) - S\left(\frac{{\cal N}-n^3}{2}\right) =
16\pi^2\beta \frac{m^2}{({\cal N}-n^3)^2}
%+8 a^2n^3\pi^2 \frac{1}{({\cal N}-n^3)^3}
%+(32a^4\pi^2-4a^2\pi^4)\frac{1}{({\cal N}-n^3)^4}
+{\cal O}\left(1/({\cal N}-n^3)^3 \right) \ .
\label{diff_action}
\eeq
The difference in the block sizes $n^1-n^2=2m$
is also given as (\ref{rel_nanbqaqb}).
Thus, (\ref{diff_action}) becomes
\beq
\Delta S \simeq  16\pi^2\beta r^2 \frac{(q^1-q^2)^2}{({\cal N}-n^3)^2} \ .
\eeq
Therefore, within the configurations with a restricted number of blocks,
the topological configurations appear in the continuum limit,
since the continuum limit is taken by
sending $\beta$ and ${\cal N}$ to infinity
with $\beta/{\cal N}$ fixed \cite{Bietenholz:2002ch}.

This situation agrees with the cases in gauge theories 
on the commutative spaces,
where one has 
\beq
\Delta S_{\rm com} =4\pi^2\beta \left(\frac{q}{({\cal N}-n^3)/2}\right)^2 \ ,
\eeq
which becomes $ 4\pi^2 (q/g L)^2$ in the continuum limit,
where $L=\epsilon({\cal N}-n^3)/2$ is the physical size of the torus,
and $g$ is the gauge coupling constant.
On the other hand, this is contrary to the cases in \cite{Aoki:2006sb, Aoki:2006zi},
where topologies are defined by the total matrix
on the NC torus.
There,
studies by classical actions and
Monte Carlo calculations 
gave $\Delta S \sim \beta ({\cal N}-n^3) $, or $\Delta S \sim \beta$ at best,
and topologically nontrivial configurations do not survive 
in the continuum limit \cite{Aoki:2006sb, Aoki:2006zi}.
Since we now define topologies by the blocks, not by the total matrix,
we recover the situations
close to the ordinary commutative spaces.

%%%%%%%%%%%%%%%%%%%%%%%%%%%%%%%%%%%%%%%%%%%%%%%%%%%%%%%
\section{Six-dimensional torus}
\label{sec:6dtorus}
\setcounter{equation}{0}

Extension of the configurations (\ref{2d_Vmu_exp})
to six dimensions is straightforward.
They are given as
\beqa
V_\mu &=& 
\begin{pmatrix}
 \Gamma_{1,\mu}^1 \otimes \id_{n^1_2} \otimes \id_{n^1_3} \otimes \id_{p^1}& &&\cr
& \Gamma_{1,\mu}^2 \otimes \id_{n^2_2} \otimes \id_{n^2_3}\otimes \id_{p^2} &&\cr
&& \ddots& \cr
&&& \Gamma_{1,\mu}^h \otimes \id_{n^h_2} \otimes \id_{n^h_3}\otimes \id_{p^h}
\end{pmatrix} \ , \n
V_{2+\mu} &=& 
\begin{pmatrix}
\id_{n^1_1} \otimes  \Gamma_{2,\mu}^1 \otimes \id_{n^1_3} \otimes \id_{p^1}& &&\cr
& \id_{n^2_1} \otimes \Gamma_{2,\mu}^2 \otimes \id_{n^2_3}\otimes \id_{p^2} &&\cr
&& \ddots& \cr
&&& \id_{n^h_1} \otimes \Gamma_{2,\mu}^h \otimes \id_{n^h_3}\otimes \id_{p^h}
\end{pmatrix} \ , \n
V_{4+\mu} &=&
\begin{pmatrix}
\id_{n^1_1} \otimes \id_{n^1_2} \otimes \Gamma_{3,\mu}^1  \otimes \id_{p^1}& &&\cr
& \id_{n^2_1} \otimes \id_{n^2_2} \otimes \Gamma_{3,\mu}^2 \otimes \id_{p^2} &&\cr
&& \ddots& \cr
&&& \id_{n^h_1} \otimes \id_{n^h_2} \otimes \Gamma_{3,\mu}^h \otimes \id_{p^h}
\end{pmatrix} \ , \n
\label{6d_Vmu_exp} 
\eeqa
with $\mu=1,2$.
In $\Gamma_{l,\mu}^a$, $n^a_l$, and $p^a$,
$a=1,\ldots,h$ specifies the blocks, and
$l=1,2,3$ specifies $T^2$'s in $T^6=T^2\times T^2\times T^2$.

The operators $\Gamma_{l,\mu}^a$ are shift operators on the dual tori
specified by a set of integers $n^a_l,m^a_l,j^a_l,k'^a_l$,
while the original tori are specified by 
$N_l, s_l, r_l, k_l$.
The integers satisfy the Diophantine equations,
\beqa
&&m^a_l j^a_l+n^a_l k'^a_l=1 \ , \\
&&2r_l s_l-k_l N_l =-1 \ ,
\eeqa
for each $a=1,\ldots,h$ and $l=1,2,3$.
The dual tori and the original tori are related by integers $q^a_l$ as
\beq
m^a_l = -s_l + k_l q^a_l \ ,~~~
n^a_l = N_l - 2r_l q^a_l \ ,
\label{rel_mn_1q_6d}
\eeq
for each $a$ and $l$.
Equation (\ref{rel_mn_1q_6d}) can be inverted as
\beq
1=2r_l m^a_l + k_l n^a_l \ ,~~~
q^a_l = N_l m^a_l + s_l n^a_l \ .
\label{rel_1q_mn_6d}
\eeq

Explicit forms of the coordinate and the shift operators on the dual tori
are given, for instance, as
\beqa
Z_{l,1}^{a} = W_{n^a_l}  ~ &,& ~~  Z_{l,2}^{a}= (V_{n^a_l})^{j^a_l} \ , \n
\Gamma_{l,1}^{a} = V_{n^a_l} ~ &,& ~~ \Gamma_{l,2}^{a} = (W_{n^a_l})^{-m^a_l} \ ,
\eeqa
in terms of the shift and  clock matrices
 (\ref{def_shit_clock}).
As shown in \cite{Aoki:2006sb}
, the configurations (\ref{6d_Vmu_exp})
are classical solutions for the action (\ref{TEK-action}).
Note also that (\ref{6d_Vmu_exp}) represents
configurations with magnetic flux in each $T^2$, and
does not exhaust 
all the topological configurations in $T^6$.

The index for the block $\psi^{ab}$ (\ref{ITprojected})
should become
\beq
\frac{1}{2}\CTr[P^{aL}P^{aR}(\gamma+\hat{\gamma})] 
=p^ap^b \prod_{l=1}^3(q^a_l-q^b_l) \ .
\label{6d_index_q_ab}
\eeq
This can also be checked as in appendix A.
Since numerical calculations take a much longer time for the six-dimensional case,
we will report on it in a future publication.

%%%%%%%%%%%%%%%%%%%%%%%%%%%%%%%%%%%%%%%%%%%%%%%%%%%%
\section{A standard model embedding in IIB matrix model}
\label{sec:SMembedding}
\setcounter{equation}{0}

We now present an example of configuration (\ref{6d_Vmu_exp})
which, when considered in the extra dimensions in the IIB matrix model,
gives matter content close to the standard model.
We can consider the situations where all the ten dimensions are compactified 
to a torus, but with an asymmetry of the sizes between our four-dimensional 
spacetime and the extra six-dimensional space.
Alternatively, we can consider the cases where our four-dimensional spacetime
is not compactified and described by Hermitian matrices as in the original 
IIB matrix model.
In this case, we consider the backgrounds as
\beqa
A_\mu &=& x_\mu \otimes \id ~~~(\mu=7,\ldots,10)~, \n
V'_\mu &=& \id \otimes V_\mu ~~~(\mu=1,\ldots,6)~,
\eeqa
with $V_\mu$ given by (\ref{6d_Vmu_exp}).
Our spacetime is represented by the backgrounds $x_\mu$.
Here, we denote our spacetime directions as $\mu=7,\ldots,10$
in order to follow the notations in the previous sections.

Let us now focus on $V_\mu$ given in (\ref{6d_Vmu_exp}).
The number of blocks is taken to be $h=4$.
The integers $q^a_l$ are taken, for instance, as 
\beq
q^{ab}_1 =
\begin{pmatrix}
0&1&0&1 \cr
&0&-1&0\cr
&&0&1\cr
&&&0
\end{pmatrix} ~,~~
q^{ab}_2 =
\begin{pmatrix}
0&1&0&3 \cr
&0&-1&2\cr
&&0&3\cr
&&&0
\end{pmatrix} ~,~~
q^{ab}_3 =
\begin{pmatrix}
0&-3&0&1 \cr
&0&3&4\cr
&&0&1\cr
&&&0
\end{pmatrix} ~,
\eeq
where we presented $q^{ab}_l=q^a_l-q^b_l$.
The lower triangle part is obtained from the upper one
by the relation $q^{ab}_l = -q^{ba}_l$.
Hence, $q^{ab}=\prod_{l=1}^3 q^{ab}_l$ becomes
\beq
q^{ab}=
\begin{pmatrix}
0&-3&0&3 \cr
&0&3&0\cr
&&0&3\cr
&&&0
\end{pmatrix} \ .
\label{q_ab_SM}
\eeq
The generation number three is obtained, 
as we will explain in detail below.

We next incorporate the gauge group structure 
by specifying the integers $p^a$ as\footnote{
Similar configurations were studied in 
\cite{Grosse:2010zq}.}
\beq
V_\mu = 
\begin{pmatrix}
 \Gamma_{\mu}^1 \otimes \id_{3}& &&\cr
& \Gamma_{\mu}^2 \otimes \id_{2} &&\cr
&& \Gamma_{\mu}^3 & \cr
&&& \Gamma_{\mu}^4 \otimes \sigma_3
\end{pmatrix} \ ,
\label{V_mu_SM}
\eeq
with $\mu =1,\ldots,6$.
$\sigma_3$ is the Pauli matrix.
The gauge group given by this background is 
$U(3) \times U(2) \times U(1)^3 \simeq SU(3) \times SU(2) \times U(1)^5$.

The fermionic matter content of the standard model is obtained 
from the fermionic matrix $\psi$ as 
\beq
\psi = 
\begin{pmatrix}
0&q&0&ud \cr
&0&\bar{l}&0\cr
&&0&\nu e \cr
&&&0
\end{pmatrix} \ ,
\label{fermion_SM}
\eeq
where each block $\psi^{ab}$ is 
$n_1^a n_2^a n_3^a p^a \times n_1^b n_2^b n_3^b p^b$ matrices.
Here, $q$ denotes the quark doublets,
$l$ the lepton doublets,
$ud$ the quark singlets,
and $\nu e$ the lepton singlets.
They are in the correct representations under the gauge group 
$SU(3) \times SU(2)$.
From (\ref{q_ab_SM}), they all have $q^{ab}$ three.
Using (\ref{6d_index_q_ab}), we find that 
they have appropriate indices that give generation number three.
The other blocks in (\ref{fermion_SM}) denoted as $0$
have a vanishing index
and do not give massless particles on our spacetime.

The hypercharge $Y$ is given by a linear combination of 
five $U(1)$ charges presented below (\ref{V_mu_SM}) as
\beq
Y=\sum_{i=1}^5 x^i Q^i \ ,
\eeq
where $Q^i=\pm 1$ with $i=1,\ldots ,5$ is the charge of 
$i$th $U(1)$ gauge group.
From the hypercharge of $q$, $u$, $d$, $l$, $\nu$, and $e$,
the following constraints are obtained:
\beqa
x^1-x^2 = 1/6 ~,~~ 
x^1-x^4 = 2/3 ~,~~ 
x^1-x^5 = -1/3~, \n
-x^2+x^3=-1/2~,~~
x^3-x^4 = 0~,~~
x^3-x^5 = -1~.
\label{hyperchargeeq}
\eeqa
Their general solutions are given by
\beq
x^1=1/6+c~,~~x^2=c~,~~x^3=x^4=-1/2+c~,~~x^5=1/2+c~,
\label{hyperchargesol}
\eeq
with $c$ being an arbitrary constant.
Since eqs. (\ref{hyperchargeeq}) depend only on the differences of $x^i$,
the solution (\ref{hyperchargesol}) is determined with an arbitrary constant shift. 
The existnece of solution is not automatically ensured,
since the number of independent variables is four
while the number of equations is six.

As the other $U(1)$ charges, 
baryon number $B$, lepton number $L$,
right-handed charge $Q_R$ and left-handed charge $Q_L$
can be considered.
Their charge for $q$, $u$, $d$, $l$, $\nu$ and $e$, 
and the corresponding values for $x^i$ are given as follows.
\beq
\begin{array}{c||c|c|c|c|c|c||c|c|c|c|c}
&q&u&d&l&\nu&e&x^1&x^2&x^3&x^4&x^5 \\ \hline\hline
Y&1/6&2/3&-1/3&-1/2&0&-1&1/6&0&-1/2&-1/2&1/2 \\ \hline
B&1/3&1/3&1/3&0&0&0&1/3&0&0&0&0 \\ \hline
L&0&0&0&1&1&1&0&0&1&0&0 \\ \hline
Q_R&0&1&1&0&1&1&0&0&0&-1&-1 \\ \hline
Q_L&1&0&0&1&0&0&0&-1&0&0&0 
\end{array}
\eeq
A linear combination of these five $U(1)$ charges
gives an overall $U(1)$ 
and does not couple to the matter.
%Only four $U(1)$ charges couple to the matter.

%%%%%%%%%%%%%%%%%%%%%%%%%%%%%%%%%%%%%%%%%%%%%%%%%%%%
\section{Conclusions and discussion}
\label{sec:conclusion}
\setcounter{equation}{0}

In this paper, 
we first introduced block-diagonal matrices for topologically nontrivial 
gauge field configurations
on a NC torus,
and found that off-diagonal blocks of the adjoint matter
can have nonzero Dirac indices.
We then showed that, by considering these configurations 
in the extra dimensions in the IIB matrix model,
chiral fermions and matter content close to the standard model
can be obtained on our four-dimensional spacetime.
In particular, generation number three was given by 
the Dirac index on the torus.
Several things remain to be clarified, 
some of which we list below.
We will report on these issues in future publications.

Our model close to the standard model gave five $U(1)$ gauge fields.
The hypercharge $U_Y(1)$ will remain massless,
while the others become massive by some dynamics of the matrix model,
or of the field theories that arise as low-energy effective theories of the matrix model.
While we did not discuss the Higgs field in the present paper,
it should be introduced, and 
the mechanism of electroweak symmetry breaking and values of the Yukawa couplings
should also be studied.

Our model is reminiscent of the intersecting D-brane models
\cite{Ibanez:2001nd,Blumenhagen:2006ci}.
There, one can obtain four-dimensional chiral fermions
by the same reason as ours, that is,
one has no remainder dimensions
normal to all the D-branes intersecting with one another
\cite{Berkooz:1996km}.
The model in \cite{Ibanez:2001nd} gives the standard model 
matter content.
Since that setting is related to ours by the T-duality, 
it is interesting to compare them with each other.
These studies may advance both string theories and matrix models.

In this paper, we studied the dynamics of the configurations by investigating the 
classical actions in the two-dimensional case,
and found that topologically nontrivial configurations 
appear in the continuum limit,
within the configurations with restricted number of blocks,
as in the commutative theories.
This shows a contrast to the cases in \cite{Aoki:2006sb, Aoki:2006zi},
where topologies were defined by the total matrix, not by the blocks,
and only the topologically trivial sector survives in the continuum limit.
For studying higher-dimensional cases, however,
quantum corrections become relevant and should be taken into account.
Owing to the quantum corrections with the noncommutativity of the torus,
a topologically nontrivial sector may arise
with higher probability than the trivial sector,
as shown in \cite{Aoki:2006zi}.
Then, the generation number three might be chosen dynamically.

We hope to study the dynamics over wider regions in the configuration space, 
including various compactifications,
in the IIB matrix model.
From these studies,
we might be able to find that
the standard model or its extension is obtained as a unique solution from
the IIB matrix model or its variants.
Or,
more complicated structures of the vacuum, such as the landscape \cite{Susskind:2003kw}, 
might be found.
Even in this case, since the matrix model has
the definite measure as well as the action,
we can define probabilities taking account of the measure,
and discuss entropy on the landscape.
The matrix models make these studies possible.

\appendix

%%%%%%%%%%%%%%%%%%%%%%%%%%%%%%%%%%%%%%%%%%%%%%%%%%%%%%%%%%
\section{Calculations of the Index}
\label{sec:cal_index}
\setcounter{equation}{0}

In this appendix, we calculate the index of the Dirac operator
for the backgrounds (\ref{2d_Vmu_exp})
and confirm that eq.~(\ref{2d_index_q_ab}) is indeed satisfied.
It is sufficient to consider the case with $h=2$ and
$p^1=p^2=1$.
For the off-diagonal block $\psi^{12}$ of the matter field $\psi$,
the operation $V_\mu \psi V_\mu^\dag$ becomes
$\Gamma^1_\mu \psi^{12} \Gamma_\mu^{2\dag}$.
Hereafter, we will write $\psi^{12}$ simply as $\psi$.
By using the explicit forms of $\Gamma^a_\mu$ in (\ref{2d_ZG_exp}),
we obtain
\beqa
(\Gamma^1_1 \psi \Gamma_1^{2\dag})_{i,j} &=& \psi_{i+1,j+1} \ , \n
(\Gamma^1_2 \psi \Gamma_2^{2\dag})_{i,j} &=& 
(\omega_{n^1})^{-m^1(i-1)} (\omega_{n^2})^{m^2(j-1)}\psi_{i,j} \ ,
\label{trans_12}
\eeqa
with $\omega_n = \e^{2 \pi i/n}$.
Here, $\psi_{ij}$ represents $ij$ components of the matrix $\psi$.

The matrix $\psi$ is $n^1 \times n^2$,
and (\ref{trans_12}) is invariant under 
identifications $i \sim i+n^1$ and $j \sim j+n^2$.
When $n^1$ and $n^2$ are coprime,
$\psi_{i,j}$ with $i=1,\ldots ,n^1$ and $j=1,\ldots ,n^2$
are mapped one-to-one by the above identifications
to $\psi_{i,i}$ with $i=1, \ldots ,n^1n^2$,
which we denote as $\psi_i$:
\beq
\psi_{i,j} \sim \psi_{i,i} \equiv \psi_i \ .
\eeq
Then, (\ref{trans_12}) is rewritten as 
\beqa
(\Gamma^1_1 \psi \Gamma_1^{2\dag})_i &=& \psi_{i+1}  \ , \n
(\Gamma^1_2 \psi \Gamma_2^{2\dag})_{i} &=&
(\omega_{n^1n^2})^{-q^{12}(i-1)} \ ,
\eeqa
with $q^{12}=q^1-q^2$.
In the second equation, we used the relation (\ref{qab_nm_qaqb}).
$\Gamma^{1\dag}_1 \psi \Gamma_1^{2}$ and
$\Gamma^{1\dag}_2 \psi \Gamma_2^{2}$ are similarly estimated.
It then follows from (\ref{def-cov-shift}) that
\beqa
\epsilon((\nabla_1^*+\nabla_1)\psi)_i 
&=&  \psi_{i+1} - \psi_{i-1} \ , \n
\epsilon ((\nabla_2^*+\nabla_2)\psi)_i 
&=& - 2 i \sin\left(\frac{2\pi}{n^1n^2}q^{12}(i-1)\right) \psi_i \ , \n
\epsilon^2 (\nabla_1^*\nabla_1\psi)_i 
&=&  \psi_{i+1} -2\psi_i + \psi_{i-1} \ , \n
\epsilon^2 (\nabla_2^*\nabla_2\psi)_i 
&=& 2\left[\cos\left(\frac{2\pi}{n^1n^2}q^{12}(i-1)\right) -1\right] \psi_i \ .
\label{nabla_2d_exp}
\eeqa

The operator H in (\ref{H-def}) is written as
\beq
H=
\begin{pmatrix}
1+\frac{\epsilon^2}{2}(\nabla_1^*\nabla_1+\nabla_2^*\nabla_2)&
-\frac{\epsilon}{2}(\nabla_1^*+\nabla_1)+i \frac{\epsilon}{2}(\nabla_2^*+\nabla_2) \cr
\frac{\epsilon}{2}(\nabla_1^*+\nabla_1)+i \frac{\epsilon}{2}(\nabla_2^*+\nabla_2)&
-1-\frac{\epsilon^2}{2}(\nabla_1^*\nabla_1+\nabla_2^*\nabla_2)
\end{pmatrix}
\label{H_2d_exp}
\eeq
by taking $\gamma_\mu = \sigma_\mu$ for $\mu=1,2$
and $\gamma =\sigma_3$.
Equations (\ref{nabla_2d_exp}) and (\ref{H_2d_exp}) give the explicit
operation of $H$ on $\psi_{i,\alpha}$, where
$\alpha=1,2$ is spinor index.
In particular, 
the operator $H$ depends only on the two integers $n^1 n^2$ and $q^{12}$.

The index of the GW Dirac operator is given by the difference
in the numbers of the positive and negative eigenvalues of the operator $H$.
We thus diagonalized it numerically.
In figure \ref{fig:index},
we plot the indices for various values of $q^{12}$ with $n^1 n^2$ fixed.
The result is periodic in $q^{12}$ with periodicity $n^1 n^2$,
and asymmetric under an exchange of $q^{12}$ to $-q^{12}$.
The graphs have similar forms irrespective of the values of $n^1 n^2$.
For $n^1 n^2=399$, which is presented in the left figure,
we find that the index takes the identical value with $q^{12}$,
and thus, eq.~(\ref{2d_index_q_ab}) is satisfied,
in the region $|q^{12}| \le 113$.
For $n^1 n^2=1295$, it is satisfied in the region
$|q^{12}| \le 367$.
\begin{figure}
\includegraphics[height=4cm]{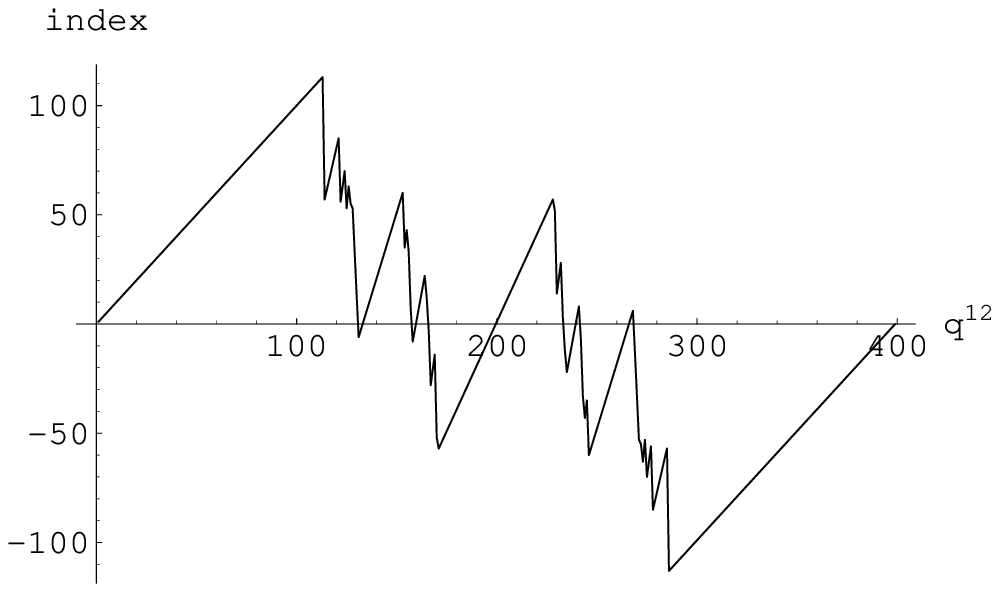}
\includegraphics[height=4cm]{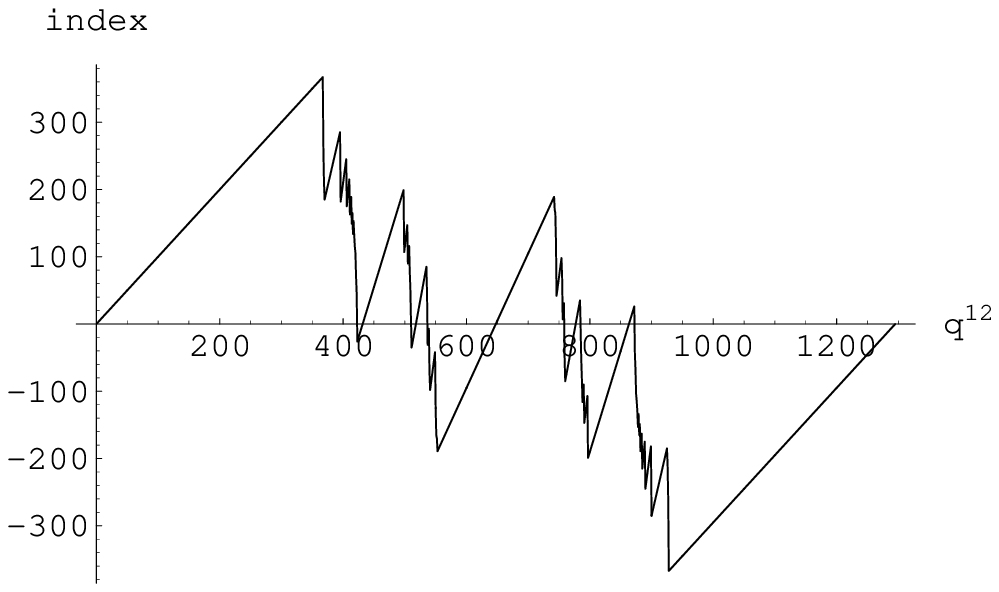}
 \caption{The indices are plotted for various values of $q^{12}$ with $n^1 n^2$ fixed.
 On the left, we take $n^1 n^2=399$, while on the right, we take $n^1 n^2=1295$.}
\label{fig:index}
\end{figure}

In figure \ref{fig:nqindex}, we plot the values of $n^1 n^2$ and $q^{12}$,
where eq.~(\ref{2d_index_q_ab}) is not satisfied.
Because of the periodicity in $q^{12}$, it is enough to survey the region
$-(n^1 n^2-1)/2 \le q^{12} \le (n^1 n^2-1)/2$ for odd $n^1 n^2$,
and $-n^1 n^2/2 +1 \le q^{12} \le n^1 n^2/2$ for even $n^1 n^2$.
From the left figure, we find that, within $n^1 n^2 \le 21$,
eq.~(\ref{2d_index_q_ab}) is satisfied at least in the 
region $|q^{12}| < (2/7) n^1 n^2$.
For $n^1 n^2 \le 101$,
which is presented in the right figure,
such safety region that ensures (\ref{2d_index_q_ab}) becomes 
$|q^{12}| < (23/81) n^1 n^2$.
For $n^1 n^2 \le 201$,
it becomes $|q^{12}| < (44/155) n^1 n^2$.
For $n^1 n^2 \le 501$,
it becomes $|q^{12}| < (128/451) n^1 n^2$.
The coefficients $2/7, 23/81, 44/155, 128/451$ slightly 
decrease as we increase $n^1n^2$.
They actually take
\beq
\frac{(22+1)l +(20+1) m}{(77+4) l + (70+4) m}
\label{81l74m}
\eeq
with $l=1$ and $m=0,1,\ldots ,24$
up to $n^1 n^2=1857$ \footnote{
The pattern (\ref{81l74m}) further continues as
with $l=2$ and $m=24,25,\ldots $,
although the safety region
does not change unless
$m$ goes beyond $48$.
We have checked this pattern until $m=45$, that is, $n^1 n^2=3492$.},
and thus, they are bounded from below by $21/74$.
We then conclude that,
for any values of $n^1 n^2$, 
eq.~(\ref{2d_index_q_ab}) is satisfied at least 
in the region $|q^{12}| < (1/3.53) n^1 n^2$.
\begin{figure}
\includegraphics[height=4cm]{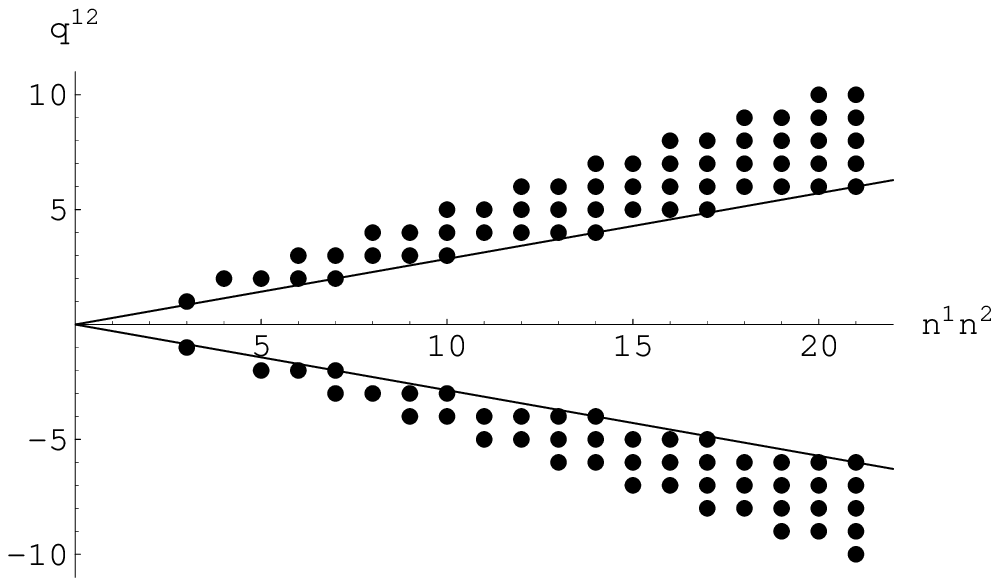}
\includegraphics[height=4cm]{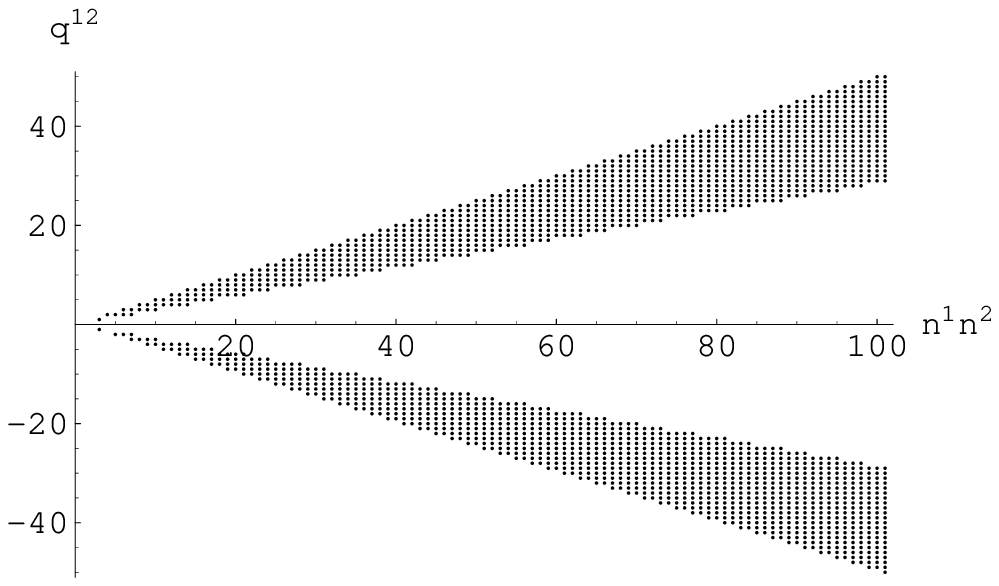}
 \caption{The values of $n^1 n^2$ and $q^{12}$,
where eq.~(\ref{2d_index_q_ab}) is not satisfied, are plotted.
Because of the periodicity in $q^{12}$, we survey the region
$-(n^1 n^2-1)/2 \le q^{12} \le (n^1 n^2-1)/2$ for odd $n^1 n^2$,
and $-n^1 n^2/2 +1 \le q^{12} \le n^1 n^2/2$ for even $n^1 n^2$.
On the left, the region $3 \le n^1 n^2 \le 21$ is shown, while on the right,
the region $3 \le n^1 n^2 \le 101$ is shown.
The lines in the left figure represent
$q^{12}=\pm (2/7) n^1 n^2$. }
\label{fig:nqindex}
\end{figure}

In fact, from the constraint (\ref{rel_nanbqaqb}), 
$n^1 n^2$ and $q^{12}$ are required to satisfy
\beq
n^1 n^2 = 2|rq^{12}|  n + (n)^2 \ ,
\label{const_n1n2q12}
\eeq
for some positive integer $n$.
Then, only the cases with $|r|=1$ and $n=1$,
which give $n^1 n^2=2|q^{12}|+1$,
are really allowed in the dotted region in figure \ref{fig:nqindex},
where eq.~(\ref{2d_index_q_ab}) is not satisfied.
They correspond to the highest and lowest points for odd $n^1 n^2$
in figure \ref{fig:nqindex}.
We therefore find that eq.~(\ref{2d_index_q_ab}) is satisfied in general,
except for the rare cases with $|r|=1$,
$n^1=1$, and $n^2=2|q^{12}|+1$,
or the cases with $n^1$ and $n^2$ reversed.


\begin{thebibliography}{99}
\bibitem{Banks:1996vh}
T.~Banks, W.~Fischler, S.~H.~Shenker and L.~Susskind,
%``M theory as a matrix model: A conjecture,''
Phys.\ Rev.\ D {\bf 55}, 5112 (1997)
[arXiv:hep-th/9610043].
%%CITATION = HEP-TH 9610043;%%

\bibitem{IKKT}
N.~Ishibashi, H.~Kawai, Y.~Kitazawa and A.~Tsuchiya,
%``A large-N reduced model as superstring,''
Nucl.\ Phys.\ B {\bf 498}, 467 (1997)
[arXiv:hep-th/9612115].
%%CITATION = HEP-TH 9612115;%%
%
For a review:
H.~Aoki, S.~Iso, H.~Kawai, Y.~Kitazawa, A.~Tsuchiya and T.~Tada,
%``IIB matrix model,''
Prog.\ Theor.\ Phys.\ Suppl.\  {\bf 134}, 47 (1999)
[arXiv:hep-th/9908038].
%%CITATION = HEP-TH 9908038;%%

%\cite{Aoki:1998vn}
\bibitem{Aoki:1998vn}
  H.~Aoki, S.~Iso, H.~Kawai, Y.~Kitazawa and T.~Tada,
  %``Space-time structures from IIB matrix model,''
  Prog.\ Theor.\ Phys.\  {\bf 99}, 713 (1998)
  [arXiv:hep-th/9802085].
  %%CITATION = PTPKA,99,713;%%

%\cite{Nishimura:2001sx}
\bibitem{Nishimura:2001sx}
  J.~Nishimura and F.~Sugino,
  %``Dynamical generation of four-dimensional space-time in the IIB matrix
  %model,''
  JHEP {\bf 0205}, 001 (2002)
  [arXiv:hep-th/0111102];
  %%CITATION = JHEPA,0205,001;%%
%\cite{Kawai:2002jk}
%\bibitem{Kawai:2002jk}
  H.~Kawai, S.~Kawamoto, T.~Kuroki, T.~Matsuo and S.~Shinohara,
  %``Mean field approximation of IIB matrix model and emergence of four
  %dimensional space-time,''
  Nucl.\ Phys.\  B {\bf 647}, 153 (2002)
  [arXiv:hep-th/0204240].
  %%CITATION = NUPHA,B647,153;%%

\bibitem{AIMN}
H.~Aoki, S.~Iso, T.~Maeda and K.~Nagao,
  %``Dynamical generation of a nontrivial index on the fuzzy 2-sphere,''
  Phys.\ Rev.\ D {\bf 71}, 045017 (2005)
  %[Erratum-ibid.\ D {\bf 71}, 069905 (2005)]
  [arXiv:hep-th/0412052].

%\cite{Steinacker:2007ay}
\bibitem{Steinacker:2007ay}
%\cite{Aschieri:2006uw}
%\bibitem{Aschieri:2006uw}
  P.~Aschieri, T.~Grammatikopoulos, H.~Steinacker and G.~Zoupanos,
  %``Dynamical generation of fuzzy extra dimensions, 
  %dimensional reduction and
  %symmetry breaking,''
  JHEP {\bf 0609}, 026 (2006)
  [arXiv:hep-th/0606021];
  %%CITATION = JHEPA,0609,026;%%
%
  H.~Steinacker and G.~Zoupanos,
  %``Fermions on spontaneously generated spherical extra dimensions,''
  JHEP {\bf 0709}, 017 (2007)
  [arXiv:0706.0398 [hep-th]];
  %%CITATION = JHEPA,0709,017;%%
%
%\cite{Chatzistavrakidis:2009ix}
%\bibitem{Chatzistavrakidis:2009ix}
  A.~Chatzistavrakidis, H.~Steinacker and G.~Zoupanos,
  %``On the fermion spectrum of spontaneously generated fuzzy extra dimensions
  %with fluxes,''
  Fortsch.\ Phys.\  {\bf 58}, 537 (2010)
  [arXiv:0909.5559 [hep-th]].
  %%CITATION = FPYKA,58,537;%%

%\cite{Atiyah:1971rm}
\bibitem{Atiyah:1971rm}
  M.~F.~Atiyah and I.~M.~Singer,
  %``The Index Of Elliptic Operators. 5,''
  Annals Math.\  {\bf 93}, 139 (1971).
  %%CITATION = ANMAA,93,139;%%
 
\bibitem{balagovi}
A.~P.~Balachandran, T.~R.~Govindarajan and B.~Ydri,
%``The fermion doubling problem and noncommutative geometry,''
Mod.\ Phys.\ Lett.\ A {\bf 15}, 1279 (2000)
[arXiv:hep-th/9911087];
%%CITATION = HEP-TH 9911087;%%
% 
%\bibitem{balaGW}
%A.~P.~Balachandran, T.~R.~Govindarajan and B.~Ydri,
%``The fermion doubling problem and noncommutative geometry. II,''
arXiv:hep-th/0006216.
%%CITATION = HEP-TH 0006216;%% 

\bibitem{Nishimura:2001dq}
J.~Nishimura and M.~A.~Vazquez-Mozo,
%``Noncommutative chiral gauge theories on the lattice with 
%manifest star-gauge invariance,''
JHEP {\bf 0108}, 033 (2001)
[arXiv:hep-th/0107110].
%%CITATION = HEP-TH 0107110;%% 

\bibitem{AIN2}
H.~Aoki, S.~Iso and K.~Nagao,
%``Ginsparg-Wilson relation, topological invariants and finite
%noncommutative geometry,''
Phys.\ Rev.\ D {\bf 67}, 085005 (2003)
[arXiv:hep-th/0209223].
%%CITATION = HEP-TH 0209223;%% 

\bibitem{GinspargWilson}
P.~H.~Ginsparg and K.~G.~Wilson,
%``A Remnant Of Chiral Symmetry On The Lattice,''
Phys.\ Rev.\ D {\bf 25}, 2649 (1982).
%%CITATION = PHRVA,D25,2649;%% 

%\bibitem{Neuberger}
H.~Neuberger,
%``Exactly massless quarks on the lattice,''
Phys.\ Lett.\ B {\bf 417}, 141 (1998)
[arXiv:hep-lat/9707022];
%%CITATION = HEP-LAT 9707022;%%
% 
%``Vector like gauge theories with almost massless fermions on the
%lattice,''
Phys.\ Rev.\ D {\bf 57}, 5417 (1998)
[arXiv:hep-lat/9710089];
%%CITATION = HEP-LAT 9710089;%%
% 
%``More about exactly massless quarks on the lattice,''
Phys.\ Lett.\ B {\bf 427}, 353 (1998)
[arXiv:hep-lat/9801031].
%%CITATION = HEP-LAT 9801031;%% 

%\bibitem{Luscher}
M.~L\"uscher,
%``Exact chiral symmetry on the lattice and the Ginsparg-Wilson relation,''
Phys.\ Lett.\ B {\bf 428}, 342 (1998)
[arXiv:hep-lat/9802011].
%%CITATION = HEP-LAT 9802011;%% 

%\bibitem{Hasenfratzindex}
P.~Hasenfratz,
%``Prospects for perfect actions,''
Nucl.\ Phys.\ Proc.\ Suppl.\  {\bf 63}, 53 (1998)
[arXiv:hep-lat/9709110];
%%CITATION = HEP-LAT 9709110;%%
% 
P.~Hasenfratz, V.~Laliena and F.~Niedermayer,
%``The index theorem in QCD with a finite cut-off,''
Phys.\ Lett.\ B {\bf 427}, 125 (1998)
[arXiv:hep-lat/9801021];
%%CITATION = HEP-LAT 9801021;%% 
%
%\bibitem{Nieder}
F.~Niedermayer,
%``Exact chiral symmetry, topological charge and related topics,''
Nucl.\ Phys.\ Proc.\ Suppl.\  {\bf 73}, 105 (1999)
[arXiv:hep-lat/9810026].
%%CITATION = HEP-LAT 9810026;%% 

%\cite{Luscher:1981zq}
%\bibitem{Luscher:1981zq}
%  M.~L\"uscher,
  %``Topology Of Lattice Gauge Fields,''
%  Commun.\ Math.\ Phys.\  {\bf 85}, 39 (1982);
  %%CITATION = CMPHA,85,39;%%
%
%\cite{Luscher:1998du}
%\bibitem{Luscher:1998du}
  M.~Luscher,
  %``Abelian chiral gauge theories on the lattice with exact gauge
  %invariance,''
  Nucl.\ Phys.\  B {\bf 549}, 295 (1999)
  [arXiv:hep-lat/9811032].
  %%CITATION = NUPHA,B549,295;%%

%\cite{Aoki:2010hx}
\bibitem{Aoki:2010hx}
  H.~Aoki,
  %``Ginsparg-Wilson relation on a fuzzy 2-sphere for adjoint matter,''
  Phys.\ Rev.\  D {\bf 82}, 085019 (2010)
  [arXiv:1007.4420 [hep-th]].
  %%CITATION = PHRVA,D82,085019;%%

%\cite{Taylor:1996ik}
\bibitem{Taylor:1996ik}
  W.~Taylor,
  %``D-brane field theory on compact spaces,''
  Phys.\ Lett.\  B {\bf 394}, 283 (1997)
  [arXiv:hep-th/9611042].
  %%CITATION = PHLTA,B394,283;%%

%\cite{Connes:1997cr}
\bibitem{Connes:1997cr}
  A.~Connes, M.~R.~Douglas and A.~S.~Schwarz,
  %``Noncommutative geometry and matrix theory: Compactification on tori,''
  JHEP {\bf 9802}, 003 (1998)
  [arXiv:hep-th/9711162].
  %%CITATION = JHEPA,9802,003;%%

%\cite{Polychronakos:1997fw}
\bibitem{Polychronakos:1997fw}
  A.~P.~Polychronakos,
  %``Unitary matrix model for toroidal compactifications of M theory,''
  Phys.\ Lett.\  B {\bf 403}, 239 (1997)
  [arXiv:hep-th/9703073];
  %%CITATION = PHLTA,B403,239;%%
%
%\cite{Kitsunezaki:1997iu}
%\bibitem{Kitsunezaki:1997iu}
  N.~Kitsunezaki and J.~Nishimura,
  %``Unitary IIB matrix model and the dynamical generation of the space  time,''
  Nucl.\ Phys.\  B {\bf 526}, 351 (1998)
  [arXiv:hep-th/9707162];
  %%CITATION = NUPHA,B526,351;%%
%
%\cite{Tada:1999mm}
%\bibitem{Tada:1999mm}
  T.~Tada and A.~Tsuchiya,
  %``Toward a supersymmetric unitary matrix formulation of the IIB matrix
  %model,''
  Prog.\ Theor.\ Phys.\  {\bf 103}, 1069 (2000)
  [arXiv:hep-th/9903037].
  %%CITATION = PTPKA,103,1069;%%
  
%\cite{Ambjorn:1999ts}
\bibitem{Ambjorn:1999ts}
  J.~Ambjorn, Y.~M.~Makeenko, J.~Nishimura and R.~J.~Szabo,
  %``Finite N matrix models of noncommutative gauge theory,''
  JHEP {\bf 9911}, 029 (1999)
  [arXiv:hep-th/9911041];
  %%CITATION = JHEPA,9911,029;%%
%
 %\cite{Ambjorn:2000nb}
%\bibitem{Ambjorn:2000nb}
%  J.~Ambjorn, Y.~M.~Makeenko, J.~Nishimura and R.~J.~Szabo,
  %``Nonperturbative dynamics of noncommutative gauge theory,''
  Phys.\ Lett.\  B {\bf 480}, 399 (2000)
  [arXiv:hep-th/0002158];
  %%CITATION = PHLTA,B480,399;%%
%
%\cite{Ambjorn:2000cs}
%\bibitem{Ambjorn:2000cs}
%  J.~Ambjorn, Y.~M.~Makeenko, J.~Nishimura and R.~J.~Szabo,
  %``Lattice gauge fields and discrete noncommutative Yang-Mills theory,''
  JHEP {\bf 0005}, 023 (2000)
  [arXiv:hep-th/0004147].
  %%CITATION = JHEPA,0005,023;%%
  
%\cite{Aoki:2008ik}
\bibitem{Aoki:2008ik}
  H.~Aoki, J.~Nishimura and Y.~Susaki,
  %``Finite-matrix formulation of gauge theories on a non-commutative torus with
  %twisted boundary conditions,''
  JHEP {\bf 0904}, 055 (2009)
  [arXiv:0810.5234 [hep-th]].
  %%CITATION = JHEPA,0904,055;%%

%\cite{Aoki:2002jt}
\bibitem{Aoki:2002jt}
  H.~Aoki, S.~Iso and T.~Suyama,
  %``Orbifold matrix model,''
  Nucl.\ Phys.\  B {\bf 634}, 71 (2002)
  [arXiv:hep-th/0203277];
  %%CITATION = NUPHA,B634,71;%%
%
%\cite{Chatzistavrakidis:2010xi}
%\bibitem{Chatzistavrakidis:2010xi}
  A.~Chatzistavrakidis, H.~Steinacker and G.~Zoupanos,
  %``Orbifolds, fuzzy spheres and chiral fermions,''
  JHEP {\bf 1005}, 100 (2010)
  [arXiv:1002.2606 [hep-th]].
  %%CITATION = JHEPA,1005,100;%% 

%\cite{Itoyama:1997gm}
\bibitem{Itoyama:1997gm}
  H.~Itoyama and A.~Tokura,
  %``USp(2k) matrix model: F theory connection,''
  Prog.\ Theor.\ Phys.\  {\bf 99}, 129 (1998)
  [arXiv:hep-th/9708123];
  %%CITATION = PTPKA,99,129;%%
%
%\cite{Itoyama:1998et}
%\bibitem{Itoyama:1998et}
%  H.~Itoyama and A.~Tokura,
  %``USp(2k) matrix model: Nonperturbative approach to orientifolds,''
  Phys.\ Rev.\  D {\bf 58}, 026002 (1998)
  [arXiv:hep-th/9801084].
  %%CITATION = PHRVA,D58,026002;%%

%\cite{Aoki:1999vr}
\bibitem{Aoki:1999vr}
  H.~Aoki, N.~Ishibashi, S.~Iso, H.~Kawai, Y.~Kitazawa and T.~Tada,
  %``Noncommutative Yang-Mills in IIB matrix model,''
  Nucl.\ Phys.\  B {\bf 565}, 176 (2000)
  [arXiv:hep-th/9908141].
  %%CITATION = NUPHA,B565,176;%%
  
%\cite{Eguchi:1982nm}
\bibitem{Eguchi:1982nm}
  T.~Eguchi and H.~Kawai,
  %``Reduction Of Dynamical Degrees Of Freedom In The Large N Gauge Theory,''
  Phys.\ Rev.\ Lett.\  {\bf 48}, 1063 (1982).
  %%CITATION = PRLTA,48,1063;%%

%\cite{GonzalezArroyo:1982ub}
\bibitem{GonzalezArroyo:1982ub}
  A.~Gonzalez-Arroyo and M.~Okawa,
  %``A Twisted Model For Large N Lattice Gauge Theory,''
  Phys.\ Lett.\  B {\bf 120}, 174 (1983);
  %%CITATION = PHLTA,B120,174;%%
%
%\cite{GonzalezArroyo:1982hz}
%\bibitem{GonzalezArroyo:1982hz}
%  A.~Gonzalez-Arroyo and M.~Okawa,
  %``The Twisted Eguchi-Kawai Model: A Reduced Model For Large N Lattice Gauge
  %Theory,''
  Phys.\ Rev.\  D {\bf 27}, 2397 (1983).
  %%CITATION = PHRVA,D27,2397;%%

\bibitem{Iso:2002jc}
S.~Iso and K.~Nagao,
%``Chiral anomaly and Ginsparg-Wilson relation on the noncommutative torus,''
Prog.\ Theor.\ Phys.\  {\bf 109}, 1017 (2003)
[arXiv:hep-th/0212284].
%%CITATION = HEP-TH 0212284;%% 

%\cite{Aoki:2006sb}
\bibitem{Aoki:2006sb}
  H.~Aoki, J.~Nishimura and Y.~Susaki,
  %``The index theorem in gauge theory on a discretized 2d non-commutative
  %torus,''
  JHEP {\bf 0702}, 033 (2007)
  [arXiv:hep-th/0602078].
  %%CITATION = JHEPA,0702,033;%%

%\cite{Aoki:2006zi}
\bibitem{Aoki:2006zi}
  H.~Aoki, J.~Nishimura and Y.~Susaki,
  %``Suppression of topologically nontrivial sectors in gauge theory on 2d
  %non-commutative geometry,''
  JHEP {\bf 0710}, 024 (2007)
  [arXiv:hep-th/0604093].
  %%CITATION = JHEPA,0710,024;%%

%\cite{Aoki:2009fs}
\bibitem{Aoki:2009fs}
 H.~Aoki, J.~Nishimura and Y.~Susaki,
  %``Dominance of a single topological sector in gauge theory on non-commutative
  %geometry,''
  JHEP {\bf 0909}, 084 (2009)
  [arXiv:0907.2107 [hep-th]].
  %%CITATION = JHEPA,0909,084;%%  

\bibitem{AIN3}
H.~Aoki, S.~Iso and K.~Nagao,
%``Ginsparg-Wilson relation and 't Hooft-Polyakov monopole on fuzzy 2-sphere,''
Nucl.\ Phys.\ B {\bf 684}, 162 (2004)
[arXiv:hep-th/0312199];
%%CITATION = HEP-TH 0312199;%% 
%
%\bibitem{AIM}
  H.~Aoki, S.~Iso and T.~Maeda,
  %``On the Ginsparg-Wilson Dirac operator in the monopole backgrounds on the
  %fuzzy 2-sphere,''
  Phys.\ Rev.\  D {\bf 75}, 085021 (2007)
  [arXiv:hep-th/0610125];
  %%CITATION = PHRVA,D75,085021;%%
%
%\cite{Aoki:2008qta}
%\bibitem{Aoki:2008qta}
  H.~Aoki, Y.~Hirayama and S.~Iso,
  %``Index theorem in spontaneously symmetry-broken gauge theories on fuzzy
  %2-sphere,''
  Phys.\ Rev.\  D {\bf 78}, 025028 (2008)
  [arXiv:0804.0568 [hep-th]].
  %%CITATION = PHRVA,D78,025028;%%
For a review:
%\cite{Aoki:2007zc}
%\bibitem{Aoki:2007zc}
  H.~Aoki,
  %``Index Theorem in Finite Noncommutative Geometry,''
  Prog. \ Theor. \ Phys. \ Suppl. \ {\bf 171}, 228 (2007)
  [arXiv:0706.3078 [hep-th]].
  %%CITATION = ARXIV:0706.3078;%%

%\cite{Aoki:2009cv}
\bibitem{Aoki:2009cv}
  H.~Aoki, Y.~Hirayama and S.~Iso,
  %``Construction of a topological charge on fuzzy S^2 x S^2 via Ginsparg-Wilson
  %relation,''
  Phys.\ Rev.\  D {\bf 80}, 125006 (2009)
  [arXiv:0909.5252 [hep-th]].
  %%CITATION = PHRVA,D80,125006;%%  
     
%\cite{Bietenholz:2002ch}
\bibitem{Bietenholz:2002ch}
  W.~Bietenholz, F.~Hofheinz and J.~Nishimura,
  %``The renormalizability of 2D Yang-Mills theory on a non-commutative
  %geometry,''
  JHEP {\bf 0209}, 009 (2002)
  [arXiv:hep-th/0203151].
  %%CITATION = JHEPA,0209,009;%%

%\cite{Grosse:2010zq}
\bibitem{Grosse:2010zq}
  H.~Grosse, F.~Lizzi and H.~Steinacker,
  %``Noncommutative gauge theory and symmetry breaking in matrix models,''
  Phys.\ Rev.\  D {\bf 81}, 085034 (2010)
  [arXiv:1001.2703 [hep-th]].
  %%CITATION = PHRVA,D81,085034;%%   
  
   %\cite{Ibanez:2001nd}
\bibitem{Ibanez:2001nd}
  L.~E.~Ibanez, F.~Marchesano and R.~Rabadan,
  %``Getting just the standard model at intersecting branes,''
  JHEP {\bf 0111}, 002 (2001)
  [arXiv:hep-th/0105155].
  %%CITATION = JHEPA,0111,002;%%

  %\cite{Blumenhagen:2006ci}
\bibitem{Blumenhagen:2006ci}
  R.~Blumenhagen, B.~Kors, D.~Lust and S.~Stieberger,
  %``Four-dimensional String Compactifications with D-Branes, Orientifolds   and
  %Fluxes,''
  Phys.\ Rept.\  {\bf 445}, 1 (2007)
  [arXiv:hep-th/0610327].
  %%CITATION = PRPLC,445,1;%% 
 
%\cite{Berkooz:1996km}
\bibitem{Berkooz:1996km}
  M.~Berkooz, M.~R.~Douglas and R.~G.~Leigh,
  %``Branes intersecting at angles,''
  Nucl.\ Phys.\  B {\bf 480}, 265 (1996)
  [arXiv:hep-th/9606139].
  %%CITATION = NUPHA,B480,265;%%

%\cite{Susskind:2003kw}
\bibitem{Susskind:2003kw}
  L.~Susskind,
  %``The anthropic landscape of string theory,''
  arXiv:hep-th/0302219.
  %%CITATION = HEP-TH/0302219;%%
\end{thebibliography}
\end{document}